\documentclass[12pt]{article}

\usepackage{graphicx}

\newcommand{\BR}{{\rm BR}}
\newcommand{\CH}{{\rm CH}}

\newcommand{\beq}{\begin{equation}}
\newcommand{\eeq}{\end{equation}}
\newcommand{\beqa}{\begin{eqnarray}}
\newcommand{\eeqa}{\end{eqnarray}}
\newcommand{\beqar}{\begin{eqnarray*}}
\newcommand{\eeqar}{\end{eqnarray*}}

\begin{document}
\thispagestyle{empty}

\hfill{\sc UG-FT-300/12}

\vspace*{-2mm}
\hfill{\sc CAFPE-170/12}

\vspace{25pt}
\begin{center}
{\textbf{\Large Heavy neutrino decays at MiniBooNE}}

\vspace{25pt}

Manuel Masip$^a$, Pere Masjuan$^{a,b}$, Davide Meloni$^c$
\vspace{12pt}

\textit{
$^a$CAFPE and Departamento de F{\'\i}sica Te\'orica y del Cosmos}\\ 
\textit{Universidad de Granada, 18071 Granada, Spain}\\
\vspace{7pt}
\textit{$^b$Institut f\"ur Kernphysik }\\ 
\textit{Johannes Gutenberg Universit\"at, 55099  Mainz, Germany}\\
\vspace{7pt}
\textit{$^c$Dipartimento di Fisica `E.~Amaldi', 
Universit\`a di Roma Tre}\\
\textit{INFN, Sezione di Roma Tre, 00146 Rome, Italy}\\
\vspace{15pt}
\texttt{masip@ugr.es, masjuan@kph.uni-mainz.de, meloni@fis.uniroma3.it}
\end{center}

\vspace{25pt}

\date{\today}

\begin{abstract}
It has been proposed that a sterile neutrino $\nu_h$ 
with $m_h\approx 50$ MeV and a dominant decay mode 
$\nu_h\to \nu \gamma$ may be the origin
of the experimental anomaly observed at LSND. We define a
particular model that could also explain the MiniBooNE excess
consistently with the data at other neutrino experiments 
(radiative muon capture at TRIUMF, T2K, or single photon  
at NOMAD). 
The key ingredients are {\it (i)} its long lifetime 
($\tau_h\approx 3\!\!-\!\!7 \times 10^{-9}$ s), which introduces
a $1/E$ dependence with the event energy, and
{\it (ii)} its Dirac nature, which implies a photon
preferably emitted opposite to the beam direction and further 
reduces the event energy at MiniBooNE.
We show that these neutrinos are mostly produced through 
electromagnetic interactions with nuclei, and that T2K
observations force ${\rm BR}(\nu_h\to \nu_\tau\gamma)\le 0.01\approx 
{\rm BR}(\nu_h\to \nu_\mu\gamma)$. The scenario implies then the 
presence of a second sterile neutrino $\nu_{h'}$ which 
is lighter, longer lived and less mixed with the standard 
flavors than $\nu_h$.
Since such particle 
would be copiously produced in air showers through
$\nu_h\to \nu_{h'} \gamma$ decays, we comment 
on the possible {\it contamination} that its 
photon-mediated elastic interactions with matter could introduce 
in dark matter experiments.

\end{abstract}

\newpage

\section{Introduction}
Although we have today evidence that neutrinos have masses and 
mixings \cite{Conrad:2007ea}, there are still basic questions 
that past experiments have been unable to answer. The most
important one concerns their Dirac or Majorana nature: 
Are neutrino masses purely electroweak (EW) or they are revealing a 
new fundamental scale? Other questions like $CP$ violation or 
the presence of extra neutrino species often use minimality 
as guiding principle, an argument that has worked 
well in the quark sector. In this sense, the three-flavor 
framework with $\Delta m_{12}^2 \approx 7.9\times 10^{-5}
\;{\rm eV^2}$, 
$\Delta m_{23,13}^2 \approx 2.4\times 10^{-3}\;{\rm eV^2}$
and mixings $\sin^2\theta_{12}\approx 0.30$, 
$\sin^2\theta_{23}\approx 0.50$, 
$\sin^2\theta_{13}\approx 0.01$ seems to fit well 
the data from solar, atmospheric and reactor experiments.

This basic picture, however, has faced a series of 
{\it persistent} anomalies in experiments with neutrino
beams from particle accelerators. Basically, muon
neutrinos of energy below 1 GeV seem to experience 
an excess of charged-current (CC) interactions with
an electron in the final state. The interpretation of 
these events in terms of $\nu_\mu\to \nu_e$ oscillations 
would require the presence of (several) 
sterile neutrinos \cite{Abazajian:2012ys}, and it fits only marginally 
the combined results of LSND, KARMEN and MiniBooNE.

Here we will discuss a very different possibility. In particular, 
we will explore a variation of the heavy-neutrino model 
proposed in \cite{Gninenko:2010pr} to explain the LSND/KARMEN 
anomaly. LSND \cite{Athanassopoulos:1996jb} was
designed to observe the interaction of muon antineutrinos 
of $E_\nu\le 52$ MeV
after a 30 m flight. The results revealed 
an excess that was interpreted as $\bar \nu_\mu \to 
\bar \nu_e$ oscillations followed by a CC quasi-elastic
interaction giving an observable $e^+$ and a free neutron. 
The neutron would then be captured to form 
a deuteron plus an also observable 2.2 MeV photon. 
The LSND signal, however,  
was not confirmed by KARMEN \cite{Armbruster:2002mp}
using a  similar technique.

In \cite{Gninenko:2010pr} Gninenko makes a very compelling 
case for a massive neutrino 
as a possible solution to the LSND/KARMEN puzzle. 
In addition to the $\bar \nu_\mu$'s from the decay-at-rest
of  $\mu^+$ leptons, 
the LSND flux  included 60--200 MeV muon neutrinos 
from the decay-in-flight of $\pi^+$
mesons. In KARMEN, however, the location of the detector
(defining a large angle relative to the incident proton 
beam) eliminates these neutrinos, selecting only 
antineutrinos of  $E\le m_\mu/2$. A
$40$--$80$ MeV sterile neutrino $\nu_h$ would then be
above the production threshold there but not at LSND. 
Gninenko shows that if $\nu_h$ has a sizable component 
along the muon flavor
($|U_{\mu h}|^2\approx 10^{-3}$--$10^{-2}$) it could explain 
the LSND excess provided that it decays fast enough 
($\tau_h \le 10^{-8}$ s)
into a light neutrino $\nu_i$ plus a photon,
\beq
\nu_h\to \nu_i \gamma\,.
\eeq
He argues that the Cherenkov light 
from a photon converted into a $e^+e^-$ pair would 
be indistinguishable from that of an electron,
and shows that the neutron hit by the initial 
$\nu_\mu$ has the {\it right} 
recoil 
to provide the correlated 2.2 MeV photon.

In Gninenko's scenario the production of the heavy neutrino
is fixed by the mixing $U_{\mu h}$, whereas the lifetime depends
on the electromagnetic (EM) coupling $\mu_{\rm tr}$ (see next 
Section). One can then easily estimate its  effects 
in other experiments. In particular, 
$\nu_h$ will be produced at MiniBooNE via $Z$ exchange 
with the nucleons in
the detector. Gninenko finds \cite{Gninenko:2010pr} that in order to 
have an impact there its lifetime 
must be reduced to 
$\tau_h \le 10^{-9}$ s,  a decade below the maximum
value suggested by LSND.
The large $U_{\mu h}$ mixing required to explain these two
experiments, however, 
will also introduce CC transitions ($\mu\to \nu_h$) with 
experimental
implications. In \cite{McKeen:2010rx}
it is shown that the model would conflict with 
observations of 
muon capture with photon emission at 
TRIUMF \cite{Bernard:2000et}, implying a signal well
above the 30\% excess (versus the standard model value) 
deduced from the data. 

Here we propose the possibility that $\nu_h$ has a 
{\it longer} lifetime, of order $3$--$7 \times 10^{-9}$ s, 
and that its dominant production channel at MiniBooNE 
is through photon (instead of $Z$) exchange with 
matter. The main effect of an increased 
$\tau_h$ is easy to understand. At TRIUMF the target
volume is much smaller than $c\tau_h$, and the 
number of events with $\nu_h$ decaying inside the detector, 
\beq
\mu^- p \to \nu_h n \to \nu_i \gamma \, n\,,
\eeq 
is proportional to
$1/\tau_h$. Therefore, an increase in the lifetime by a factor
of $3$--$7$ respect to the value assumed in \cite{McKeen:2010rx}
will reduce the number of events by the same factor.
The larger lifetime will 
also reduce the number of events at MiniBooNE. However,
as noticed in \cite{McKeen:2010rx}, the EM couplings
necessary to mediate $\nu_h$ decay may also have an impact on its
production, and we will use them to compensate this
reduction. In addition, since the decay length $\lambda_{d}$
of the heavy neutrino grows linear with its energy,
\beq
\lambda_{d}= c\, \tau_h 
{E_h \over m_h}\, \sqrt{1-{m_h^2\over E_h^2}}\,,
\label{ldec}
\eeq
the effect of a larger $\tau_h$ will be stronger on 
high than on low-energy events at MiniBooNE. 
The energy dependence that we will obtain fits better the 
data there than
neutrino oscillations or the prompt $\nu_h$-decay hypothesis
by Gninenko.

In the next sections we define a model for the heavy neutrino
and explore its
implications at MiniBooNE. In section 4 we study the
impact of the model on T2K, which further constrains the
scenario 
and defines a {\it working} model. Finally, in section 5
we sumarize our results and 
discuss the possible implications of the scenario in
other experiments.

\section{The heavy neutrino}

We will assume that $\nu_h$ is a Dirac particle of mass 
$m_h=50$ MeV with a left-handed component slightly mixed 
with the muon neutrino:
\beqa
\nu'_{h }&=&\cos \theta \;\nu_{h } + \sin \theta \;\nu_{\mu }
\,,\nonumber \\
\nu'_{\mu }&=&-\sin \theta \;\nu_{h } + \cos \theta \;\nu_{\mu }\,, 
\eeqa
where $\sin \theta=U_{\mu h}$ and $|U_{\mu h}|^2=0.003$.
We will also assume that the heavy neutrino does not introduce any
new sources of lepton-number violation. If the light 
neutrinos $\nu_i$ ($i=e,\mu,\tau$)
are Dirac particles this assumption will not have any implications on
the $\nu_i\to \nu_h$ EM transitions. 
If they are Majorana, however, it implies a relation between the
electric and the magnetic dipole transitions, 
\beq
L_{eff}\subset {1\over 2} \, \mu_{\rm tr}^i \,
\Big( \overline \nu_h\, \sigma _{\mu \nu}
 \left(1-\gamma_5\right) \nu_i + \overline \nu_i\, \sigma _{\mu \nu}  
\left(1+\gamma_5\right) \nu_h \Big) \, \partial^\mu A^\nu
\,, 
\eeq
where  we have dropped the prime 
to indicate mass eigenstates and have taken a 
real value for the 
coupling $\mu^i_{\rm tr}$. The Lagrangian
above is then $CP$ conserving, it only breaks parity. 
We will use it
to calculate the production and the
decay (in Fig.~1) of the
heavy neutrino $\nu_h$. Since the masses of the light 
neutrinos are negligible at MiniBooNE energies, lepton
number will be preserved both in EW processes and 
in the production or the decay of 
$\nu_h$, and the  light neutrinos $\nu_i$ 
($\overline \nu_i$) will always appear with negative
(positive) helicity.

\begin{figure}[!t]
\begin{center}
\includegraphics[width=0.7\linewidth]{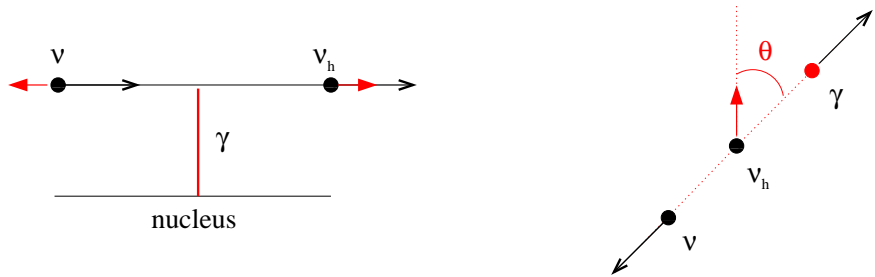} 
\end{center}
\caption{Heavy neutrino production $\nu Z \to \nu_h Z$ through
photon exchange with a nucleus (left) and decay 
$\nu_h\to \nu \gamma$  (right).
\label{fig1}} 
\end{figure}

{\bf Production cross section.} 
The production of the heavy neutrino in $\nu_\mu$ collisions 
with matter will include  $Z$- and photon-mediated processes. 
The first ones, with an amplitude proportional to $U_{\mu h}$,
will couple $\nu_\mu$ to the protons and the neutrons 
in the target oil ($\CH_2$) at MiniBooNE
(collisions with
electrons are negligible). We will show, however, that long-distance
processes mediated by the massless photon, 
with an amplitude 
proportional to  $\mu^\mu_{\rm tr}$, may be the dominant ones.

Let us consider the quasi-elastic production of $\nu_h$
through photon exchange when a neutrino of energy $E_\nu$ 
scatters off a nucleus of charge $Z$ and mass 
$M$. The initial neutrino results from
a meson or a muon decay, and it will always be polarized against the 
beam direction (i.e., it has a negative helicity). 
If the target is a spin 1/2 nucleus the cross section reads
\beq
{ {\rm d} \sigma(\nu_\mu Z \to \nu_h Z) \over {\rm d}t } =
{\alpha\, Z^2 F^2(t)\,(\mu_{\rm tr}^\mu)^2\over 2}\, {f(s,t)\over
t^2 (s-M^2)^2}
\eeq
where $s=M^2+2 E_\nu M$, $t=(p_{\nu_h}-p_\nu)^2$, $F(t)$ is the form
factor, and 
\beq
f(s,t) =  - 2 t (s - M^2 ) ( s - M^2 + t ) + 
m_h^2 t ( 2 s + t ) - m_h^4 (2 M^2 + t)
\,.
\eeq
If the target is a spin 0 nucleus (like $^{12}{\rm C}$) the cross
section can be obtained just by replacing the term 
$- m_h^4 t $ in $f(s,t)$ by $-t^2(t+m_h^2/2)$.
The Mandelstand variable $t$ is simply related to 
the recoil energy $T$ of the nucleus,
\beq
T=-{t\over 2 M}\,,
\eeq
whereas the scattering angle $\theta$ in the lab frame is 
\beq
\cos\theta = {E_\nu-T-{MT\over E_\nu}-{m_h^2\over 2 E_\nu}\over
\sqrt{E^2_\nu+T^2-2E_\nu T -m_h^2}}\,.
\eeq
Notice also that if the neutrino 
$\nu_h$ were massless, it would be 
always produced as a right-handed
particle of positive helicity, 
since the EM transition would flip the chirality. 
The final state with 
negative helicity gives a contribution
of order $(m_h/E_\nu)^2$ to 
the cross section above.
For $m_h=50$ MeV, at 700 MeV 
we find that only one out of $5\times 10^5$ 
massive neutrinos will have the spin against its momentum.
In contrast, the $\nu_h$ produced at MiniBooNE
through $Z$ exchange will have the opposite 
(negative) helicity.

\begin{figure}[!t]
\begin{center}
\includegraphics[width=0.46\linewidth]{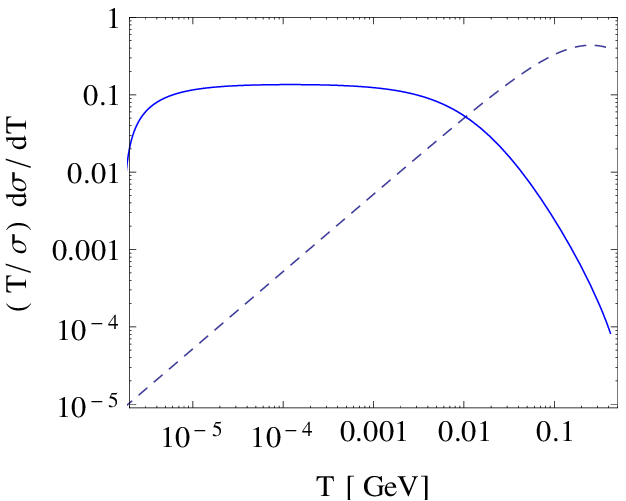} $\;\;\;$
\includegraphics[width=0.50\linewidth]{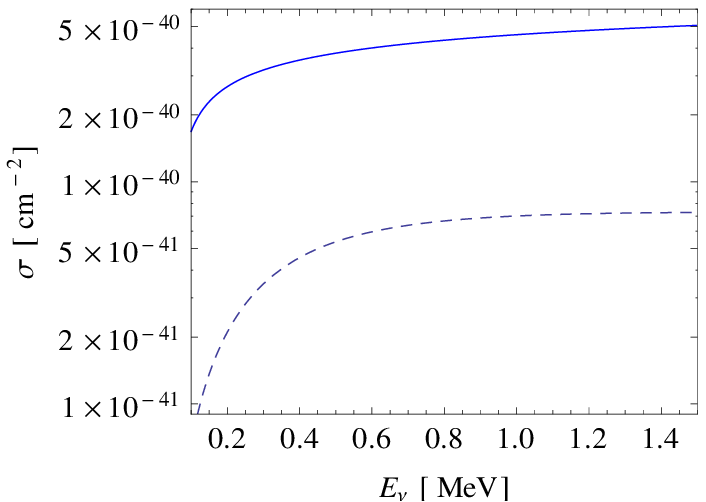} 
\end{center}
\caption{ Left: ${T\over \sigma} {{\rm d}\sigma\over 
{\rm d} T}$ for  $\nu_\mu p \to \nu_h p$ through
photon (solid) and $Z$ (dashes) exchange for $E_\nu=0.7$ GeV. 
Right: Total cross
section $\sigma( \nu_\mu\; \CH_2 \to \nu_h\; \CH_2)$ 
for $U_{\mu h}^2=0.003$  and 
$\mu_{\rm tr}^\mu=2.4\times 10^{-9}\mu_B$.
\label{fig2}} 
\end{figure}
In addition to the helicity of the final neutrino, 
the main difference between $Z$ and 
photon-mediated processes is the typical distance (or 
equivalently, the $q^2=-t$ or the recoil $T$) 
in the collision. In Fig.~2--left we plot the 
distribution ($(T/\sigma)\,{{\rm d}\sigma/{\rm d} T}$) for 
 $\nu_\mu p \to \nu_h p$ at $E=700$ MeV.
We find that the average recoil is $T=150$ MeV
when the collision is mediated by a $Z$ boson but just 2 MeV 
in processes that go through photon exchange. 

For form factor $F(t)$ describing the target we have used
the expresions in \cite{Gninenko:1998nn}. At $q^2$ between $10^{-9}$ and 
$10^{-3}$ GeV$^2$, which are dominant in these photon-mediated
collisions, 
the incident neutrino interacts coherently
with the whole nucleus. At lower values of $q^2$ electrons tend to
screen the nuclear electric charge, 
whereas at higher values the collision
is mainly with the protons inside the nucleus (we use then the usual 
EM form factors \cite{Horowitz:1993rj}). 
In Fig.~2--right we plot 
the total cross section for the quasi-elastic 
process $\nu_\mu \,\CH_2 \to \nu_h\, \CH_2$ 
for $U_{\mu h}^2=0.003$  and 
$\mu_{\rm tr}^\mu=2.4\times 10^{-9}\mu_B$ (see below).

{\bf Decay rate.} 
For the decay $\nu_h\to \nu_i\gamma$ (in Fig.~1--right), 
let us consider a neutrino $\nu_h$ at rest and 
with spin $|+\rangle$ along the $Z$ direction. It is easy to 
find \cite{Li:1981um,book1} the angular distribution of the final photon:
\beq
{ {\rm d} \Gamma \over {\rm d}\cos \theta } =
{ \mu^2_{\rm tr}\over 32 \pi }\, m_h^3\, \left( 1 - \cos \theta \right)\,
\label{distgamma}
\eeq
where $\mu^2_{\rm tr}=\sum_i (\mu^i_{\rm tr})^2$ and 
$\theta$ is the angle between the photon momentum and the $Z$ axis.
Therefore, the photon prefers to exit  against the
spin of the initial heavy neutrino. Analogously, when the particle 
decaying is an antineutrino $\bar \nu_h$ the final photon will
be more frequently emitted along the direction of the spin.

For a neutrino mass $m_h=50$ MeV, the lifetime 
$\tau_h= 5\times 10^{-9}$ s is obtained with
$\mu_{\rm tr}=7.2\times 10^{-6}\;{\rm GeV}^{-1}=
2.4\times 10^{-8}\mu_B$. This is a relatively {\it large}
value of $\mu_{\rm tr}$, as it 
will be generated at one loop and must include a 
fermion-mass insertion. The ultraviolet completion of our 
$\nu_h$ scenario would then require the presence of
extra physics at the
TeV scale (see, for example, the left-right symmetric models
in \cite{book1}).

When the  decay $\nu_h\to \nu_i \gamma$ includes 
several neutrino species ($i=\mu,\, ...$)
the branching ratios are just 
$\BR_i=(\mu^i_{\rm tr})^2/\mu_{\rm tr}^2$.
Notice also that since the beam at MiniBooNE is
composed of (mostly) muon neutrinos, the production cross section
through photon exchange is just sensitive to 
$(\mu^\mu_{\rm tr})^2$.

\section{Heavy neutrino events at MiniBooNE}

\begin{figure}[!t]
\begin{center}
\includegraphics[width=0.46\linewidth]{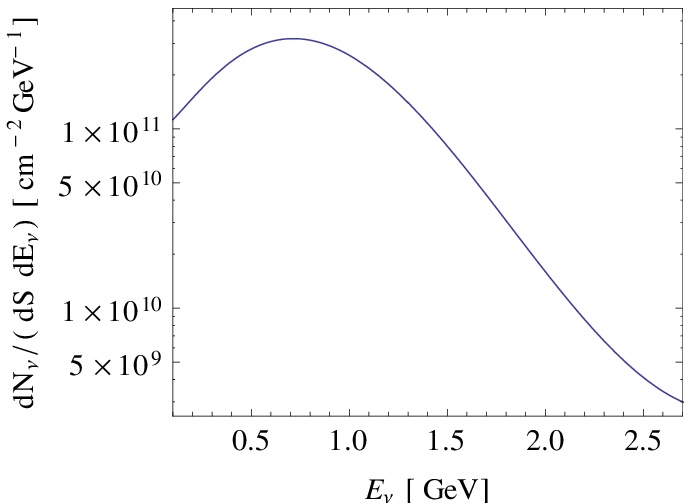} $\;\;\;$
\includegraphics[width=0.46\linewidth]{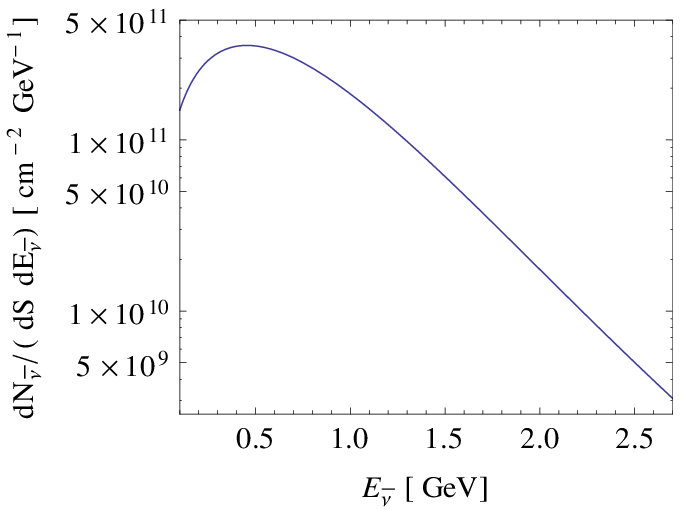} 
\end{center}
\caption{Flux at MiniBooNE in the neutrino mode for
$5.58\times 10^{20}$ POT (left) and in the antineutrino mode
for $11.27\times 10^{20}$ POT
(right).
\label{fig3}} 
\end{figure}
MiniBooNE \cite{AguilarArevalo:2008yp}
has run in neutrino \cite{AguilarArevalo:2007it}
and antineutrino \cite{AguilarArevalo:2009xn} modes. We will 
first analyze the results for neutrinos from  
$6.46\times 10^{20}$ protons on target (POT) presented in 
\cite{AguilarArevalo:2007it}.
These data were initially used by MiniBooNE 
to {\it exclude} the neutrino oscillation hypothesis 
favored by LSND (see below), although latter analyses
 emphasized the anomaly observed 
at low energies \cite{AguilarArevalo:2007it}. We will
also estimate our prediction for MiniBooNE in the
antineutrino mode with 
an exposure to $11.27\times 10^{20}$ POT
\cite{AguilarArevalo:2012va}, where the data seems more 
consistent with
the 3+1 neutrino-oscillation picture. Our fit to the total 
muon neutrino flux ${\rm d}N_\nu/({\rm d}S\, {\rm d}E_\nu)$ 
\cite{AguilarArevalo:2008yp} in these 
two cases is given in Fig.~3.

MiniBooNE is a sphere of fiducial radius $R\approx 5$ m filled up
with $\CH_2$ of density $\rho=0.86$ g/cm$^3$. When a neutrino enters the
detector it finds an oil column of length $L$, with $L$ taking values
between 0 and $2R$ depending on the point of entrance (see Fig.~4).
To define an observable $\nu_h$ event, first 
$\nu_\mu$ must interact at a distance
$l<L$ from the entrance and produce the heavy neutrino, 
and then $\nu_h$ must decay within a
distance $L-l$, {\it i.e.}, before leaving the fiducial volume. 
It is easy to
see that, while the 
probability  to interact is just
\beq 
p_{\rm i} = {\sigma \, \rho \, L\over m_{CH_2}}\,,
\eeq 
the probability to interact {\it and} decay within the distance 
$L$ becomes
\beq 
p_{\rm i+d} = {\sigma \, \rho \, L\over m_{CH_2}} 
\left( 1- {\lambda_{d}\over L} \left( 
1-e^{-{L\over \lambda_{d}}} \right) \right)\,,
\eeq
where the decay length $\lambda_{d}$, defined in
Eq.~(\ref{ldec}), depends on the 
heavy neutrino energy $E_h=E_\nu-T$ and its
lifetime $\tau_h$. Since the photon-mediated processes 
under study introduce
small recoils, in our estimate we will consider that 
$\nu_h$ is produced forward and with energy $E_h=E_\nu$.
Integrating over all the points of entrance into
the detector we obtain that the energy distribution
of heavy neutrinos decaying inside MiniBooNE is
\beq 
{{\rm d}N_{h} \over {\rm d}E_\nu} = {{\rm d}N_\nu \over 
{\rm d}S\,dE_\nu} \;
{\sigma\, V\, \rho\over m_{CH_2}}
\int_0^1{\rm d}y\, \left( 1- {\lambda_{d}\over 2 R y^{1/3}}
\left( 1-e^{-{2 R y^{1/3}\over \lambda_{d}}} \right) \right)\,,
\eeq 
\begin{figure}[!t]
\begin{center}
\includegraphics[width=0.42\linewidth]{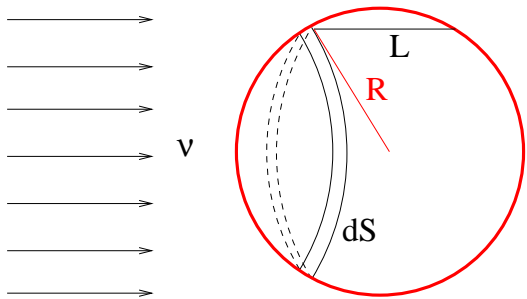} 
\end{center}
\caption{Geometry of the detector at MiniBooNE.
\label{fig4}} 
\end{figure}
where $V=4 \pi R^3/3$, 
$m_{CH_2}=(14/6.022)\times 10^{-23}$ g and the energy
dependence of the total cross section $\sigma$ and of $\lambda_d$
is understood.

The $\nu_h$ energy distribution ${\rm d}N_{h}/ {\rm d}E_\nu$ above
must then be translated into 
a distribution of visible energy.
As explained in the previous section, most of the neutrinos 
$\nu_h$ produced via photon exchange have positive helicity,
and when they decay $\nu_h \to \nu_i \gamma$ the gamma tends to
exit backwards. If $\nu_h$ carries  $E_\nu$, 
Eq.(\ref{distgamma}) implies that in the lab frame 
the energy of the 
final photon will be distributed linearly 
between $E_{\rm min}=E_\nu (1-\beta)/2$ and 
$E_{\rm max}=E_\nu (1+\beta)/2$, with 
$\beta=\sqrt{1-m_h^2/E_\nu^2}$. The distribution of $x=E_\gamma/E_\nu$
\begin{figure}[!t]
\begin{center}
\includegraphics[width=0.45\linewidth]{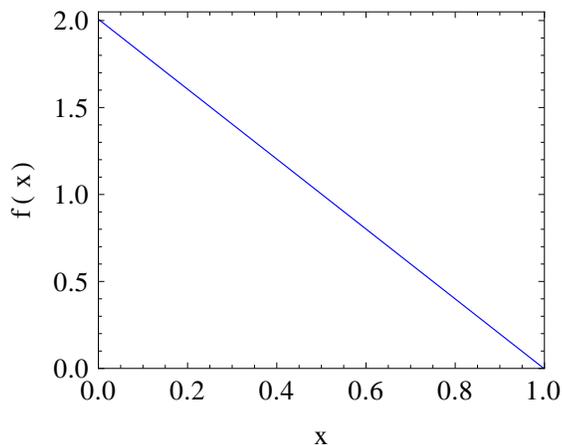} 
\end{center}
\caption{Fraction of energy taken by $\gamma$ in
$\nu_h\to \nu_i \gamma$  ($E_\nu=0.7$ GeV, $m_h=50$ MeV).
\label{fig5}} 
\end{figure}
is then just
\beq
f(x)={2\over x_{\rm max}-x_{\rm min}}-2{x -x_{\rm min}\over 
(x_{\rm max}-x_{\rm min})^2}\,,
\eeq
with $x_{\rm min(max)}=E_{\rm min(max)}/E_\nu$. 
In Fig.~5 
we plot $f(x)$ for $E_\nu=0.7$ GeV and $m_h=50$ MeV.
The energy distribution ${\rm d}N_h / {\rm d} E_\gamma$
of the photons from heavy-neutrino production and decay
is then  
obtained from the integral
\beq 
{{\rm d}N_{h} \over {\rm d}E_\gamma} = 
\int_0^1\,{\rm d}x\; {{\rm d}N_h \over {\rm d}E_\nu}(E_\gamma/x) 
\;{f(x)\over x}\,,
\eeq 
where
$dN_h /dE_\nu$ is evaluated at $E_\nu=E_\gamma/x$. 

In \cite{AguilarArevalo:2007it}  
the results are presented in terms of the 
energy of the incident neutrino, which is reconstructed from
the visible energy and 
the scattering angle of the final electron. In the
events under study here the final photon will typically 
exit forward, defining a small angle.
For example, in average a 400 MeV neutrino would
produce through photon exchange a $\nu_h$ with an 
angle of $7^o$, and
its decay would add $22^o$ to the final photon trajectory. 
\begin{figure}[!t]
\begin{center}
\includegraphics[width=0.49\linewidth]{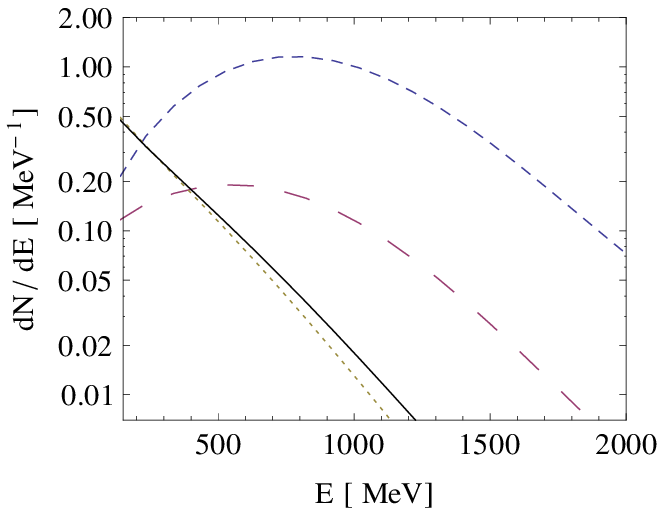}  $\;\;\;$
\includegraphics[width=0.45\linewidth]{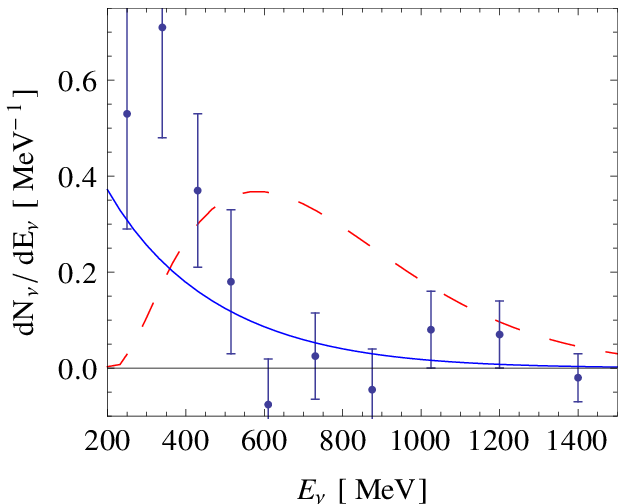}  
\end{center}
\caption{Left: Energy distribution of heavy neutrinos produced in 
the detector (dashes), of heavy neutrinos decaying inside the 
detector (long dashes), of photons from $\nu_h\to \nu_i \gamma$ (dots), 
and of 
$\nu_h$ events reconstructed as CC interactions (solid).
Right:
Energy distribution of 
$\nu_h$ events reconstructed as CC interactions (solid),
of events from neutrino oscillations for 
$\sin^2(2\theta)=0.004$ and $\Delta m^2=1\;{\rm eV}^2$ (long 
dashes), and
excess observed by MiniBooNE in the neutrino mode.
\label{fig6}} 
\end{figure}
In contrast,
a $W$-mediated process would imply an electron with an average
angle of $65^o$ (see \cite{Benhar:2006nr,Martinez:2005xe} for a study of 
quasielastic neutrino scattering).
In Fig.~6--left we plot the energy distribution ${\rm d}N/{\rm d}E$ of 
neutrinos $\nu_h$ produced inside the detector,
of neutrinos that decay inside the 
detector, of photons from $\nu_h$
decays, and of the neutrinos that 
one would reconstruct assuming that the visible energy 
comes from a CC interaction. These results
include the events where $\nu_h$ is produced through $Z$ exchange, 
which for $U_{\mu h}^2=0.003$ account for just a 12\% of the total.
We have taken  
$\tau_h= 5\times 10^{-9}$ s and $\BR_\mu=0.01$, which correspond
to $\mu^\mu_{\rm tr}=2.4\times 10^{-9}\mu_B$.

In Fig.~6--right we summarize 
our results for MiniBooNE in the neutrino mode.
 We give the energy distribution of the $\nu_h$ events 
reconstructed as $W$-mediated interactions together with
the distribution expected from
neutrino oscillations for 
$\sin^2(2\theta)=0.004$ and $\Delta m^2=1\;{\rm eV}^2$
and the excess observed at MiniBooNE. 
These oscillation parameters had been favored by
the LSND anomaly, but they imply a $500$--$700$ MeV excess
that was initially excluded by MiniBooNE.
In contrast, the long-lived heavy neutrino hypothesis seems
consistent with the data.

In Fig.~7 we plot the results from an analogous analysis 
for $11.27\times 10^{20}$ POT in the antineutrino mode 
of MiniBooNE \cite{AguilarArevalo:2012va}. The data at
$500$--$700$ MeV seem to favor the LSND oscillation
hypothesis, although the excess observed at lower energies
would still remain unexplained. Our long-lived heavy neutrino,
instead, provides a reasonable fit at all energies.

\begin{figure}[!t]
\begin{center}
\includegraphics[width=0.48\linewidth]{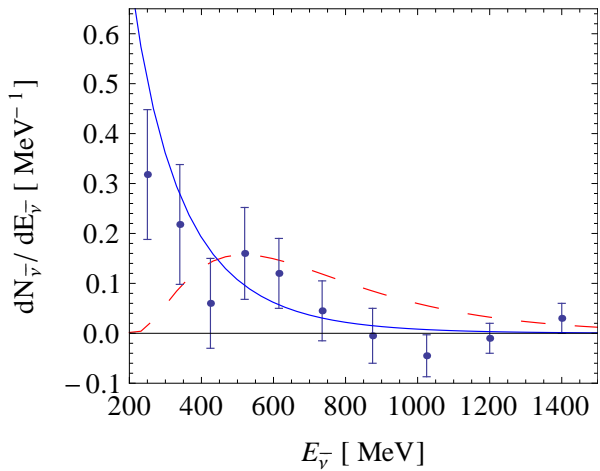}   
\end{center}
\caption{
Energy distribution of 
$\nu_h$ events reconstructed as CC $\nu$ interactions (solid),
of events from neutrino oscillations for 
$\sin^2(2\theta)=0.004$ and $\Delta m^2=1\;{\rm eV}^2$ (long dashes), and
excess observed by MiniBooNE in the antineutrino mode 
(with error bars).
\label{fig7}} 
\end{figure}

\section{T2K implications}

LSND and MiniBooNE led us to 
define a model for $\nu_h$ where the EM transition 
that fixes its lifetime, $\mu_{\rm tr}=2.4\times 10^{-8}\mu_B$,
is 10 times larger than 
the coupling required to produce it at
MiniBooNE, $\mu^\mu_{\rm tr}=2.4\times 10^{-9}\mu_B$.
This implies that the branching ratio for the decay 
into the muon-neutrino flavor 
($\nu_h\to \nu_\mu \gamma$) is
just around 1\%. 
The obvious question is then what other flavor may account for 99\%
of the decays, and the tau neutrino seems the natural candidate.

To check the consistency of this possibility one needs a $\nu_\tau$
beam, and the long-baseline experiment 
T2K \cite{Abe:2011ks,Abe:2011sj}
provides such a beam. Muon neutrinos are produced at Tokai 
with $E_\nu\approx 0.6$ GeV
and they travel $295$ km to Super-Kamiokande,
where most of them have oscillated into the tau flavor:
\beq
p_{\nu_\mu\to \nu_\tau}\approx \sin^2 2\theta_{23} \; 
\sin^2{1.27\; \Delta m_{23}^2({\rm eV}^2)\; 
L({\rm km})\over E_\nu({\rm GeV})}\,,
\eeq
with $\sin^2 2\theta_{23}\approx 1$ and 
$\Delta m_{23}^2\approx 2.4\times 10^{-3}\;{\rm eV^2}$.
Therefore, T2K would be able to measure the $\theta_{13}$ angle
through oscillations of muon
into electron neutrinos, 
\beq
p_{\nu_\mu\to \nu_e}\approx \sin^2 \theta_{23} \; \sin^2 2\theta_{13} \; 
\sin^2{1.27\; \Delta m_{23}^2({\rm eV}^2)\; 
L({\rm km})\over E_\nu({\rm GeV})}\,,
\eeq
but also the anomalous EM interactions
of  $\nu_\tau$ with matter to produce the heavy neutrino  $\nu_h$ 
\cite{Gninenko:2011hb} (see also \cite{Dib:2011hc} for 
possible effects of $\nu_h$ on tau physics).

The inner detector (ID) at Super-Kamiokande is a cylindrical tank
with a fiducial volume of radius $R=17$ m 
and 32 m high ({\it i.e.}, 22.6 tones of pure water). 
An outer detector (OD) enclosing the tank 
excludes events with activity at distances below 4.5 m
from the fiducial volume in the ID. Therefore, there are
two ways to generate a signal through $\nu_h$ decays. 
The heavy neutrino could be produced in the rock before
the detector, fly through the OD and decay 
inside the fiducial volume in the ID, or it could 
both be produced and decay inside the ID.

If only muon neutrinos can produce $\nu_h$ through photon
exchange ({\it i.e.}, 
$\mu^\mu_{\rm tr}=2.4\times 10^{-9}\mu_B$, 
$\mu^\tau_{\rm tr}=0$), 
the energy distribution of the heavy neutrinos that 
are produced 
and decay inside the ID is
\beq 
{{\rm d}N_{h} \over {\rm d}E_\nu} = {{\rm d}N_\nu \over 
{\rm d}S\,{\rm d}E_\nu} \;
{\sigma\, V\, \rho\, (1-p_{\nu_\mu\to \nu_\tau}) \over m_{H_2O}}
\int_0^{\pi/2}{\rm d}\theta\, {4\cos^2\theta\over \pi}
\left( 1- {\lambda_{d}\over 2 R \cos\theta}
\left( 1-e^{-{2 R \cos\theta \over \lambda_{d}}} \right) \right)\,,
\eeq 
where  $\lambda_{d}$ is defined in
Eq.~(\ref{ldec}), $\sigma$ is the cross section 
to produce a $\nu_h$ in the collision of $\nu_\mu$ 
with water, $\rho=1$ g/cm$^3$, and we have taken the neutrino flux 
${\rm d}N_\nu/ {\rm d}S\,{\rm d}E_\nu$  
in \cite{Abe:2011sj}.
For comparison, the distribution of $\nu_e$
producing electrons in collisions with water would just be
\beq 
{{\rm d}N_{e} \over {\rm d}E_\nu} = {{\rm d}N_\nu \over 
{\rm d}S\,{\rm d}E_\nu} \;
{\sigma_{CC}\, V\, \rho\, p_{\nu_\mu\to\nu_ e} \over m_{H_2O}}\,.
\eeq 
The calculation of the energy distribution of the neutrinos
produced in the surrounding 
rock (beyond a distance $D_0=4.5$ m of the ID)
and decaying inside the detector is also straightforward,
\beq 
{{\rm d}N'_{h} \over {\rm d}E_\nu} = {{\rm d}N_\nu \over 
{\rm d}S\,dE_\nu} \;
{\sigma\, V\, \rho_r\, (1-p_{\nu_\mu\to \nu_\tau}) \over m_{r}}\, 
{2\, \lambda_d\,  e^{-D_0 \over \lambda_d}\over \pi\,  R} \, 
\int_0^{\pi/2}{\rm d}\theta\, \cos\theta
\left( 1-e^{-{2 R \cos\theta \over \lambda_{d}}} \right)\,,
\eeq 
where we have taken 
$\rho_r=3.2$ g/cm$^3$.

We obtain that whereas $\nu_\mu\to \nu_e$ oscillations 
with $\sin^22\theta_{13}=0.1$
would imply 6 events of energy above 100 MeV, 
$\nu_\mu$ collisions producing a $\nu_h$ that 
decays into $\nu_i \gamma$
would introduce 1.1 events, 
75\% of them from heavy neutrinos created in the rock
and decaying inside the detector.
These {\it long-flying} neutrinos 
dominate over the ones produced inside the detector
due to the larger density of the soil sorrounding the
detector and the relatively
long lifetime ($\lambda_d=15$ m 
for $E_h=500$ MeV) of $\nu_h$. Therefore, in our framework
the MiniBooNE anomaly
could also
{\it contribute} to the signal observed at T2K \cite{Abe:2011sj}, 
 implying a smaller value of the mixing $\theta_{13}$. 
The contribution would be larger for a non-zero value of 
of the coupling $\mu^\tau_{\rm tr}$ that defines the
$\nu_\tau \to \nu_h$ EM transitions.
Since these
$\nu_h$ events have different kinematical and energy distributions,
it seems clear that an increased statistics there could disregard
this possibility. Incidentally, the events with $\nu_h$ produced
in the rock would be distributed preferably
near the point of entrance into the detector (the distribution
is proportional to $e^{-D/\lambda_d}$), as it has been 
observed, whereas the 
$\nu_\mu\to \nu_e$ oscillation hypothesis implies a flat distribution.

The previous result also implies that 
$\mu^\tau_{\rm tr}\le \mu^\mu_{\rm tr}$.
In particular, if the channel 
$\nu_h\to \nu_\tau \gamma$ accounted for
$99\%$ of the $\nu_h$ decays, then the number of electron-like
events at T2K from 
$\nu_\tau Z \to \nu_h Z$ EM transitions
would be around 240, well above the 6 events
that were observed. As a consequence, to be viable the model
requires an additional sterile\footnote{The electron flavor
seems also disfavored, since the 
process $\nu_e Z\to \nu_h Z$ with 
$\nu_h\to \nu_e \gamma$ in the atmosphere 
would introduce a signal 
identical to $\nu_e$ CC interactions.} neutrino $\nu_{h'}$ which
is lighter than $\nu_h$ and has the coupling
$\mu^{h'}_{\rm tr}\approx 2.4\times 10^{-8}\mu_B$ 
required to mediate $\nu_h \to \nu_{h'} \gamma$ at
the right rate ($\tau_h\approx 5\times 10^{-9}$ s).

Finally, we would like to comment on the signal 
that the model could imply at the
near detector (ND280) \cite{Abe:2011ks} in T2K.
When the neutrino beam reaches the off-axis
detector at ND280  it has not changed its muon flavor yet.
A typical event would consist of an
initial interaction producing
the heavy neutrino $\nu_h$ and 
a small energy deposition (a 10 keV--10 MeV nuclear recoil) 
in the Pi-zero detector, followed by the single photon from 
its decay at any point in the tracking system (three TPCs and
two thinner FGDs). We estimate around 2.1 events of this type per
1000 $\nu_\mu$ CC interactions, plus 0.8 events with the $\nu_h$ 
both being produced and decaying inside the Pi-zero detector.
In the tracker the photon would convert into an $e^+e^-$ pair
that could be distinguished from the single electron
plus recoil in a CC interaction, 
whereas events with neutral pions giving two 
photons seem also clearly different. 
Therefore, ND280 may 
provide some ground to probe the MiniBooNE/LSND anomaly and
establish its electron or photon origin (if any).

\section{Summary and discussion}

LSND observed an excess of $\approx 3\times 10^{-3}$ interactions
with an electron in the final state per each $\bar \nu_\mu$
CC event. The beam in this experiment included antineutrinos
of $E<60$ MeV from the decay-at-rest of $\mu^+$ leptons 
but also neutrinos of up to 200 MeV. 
KARMEN tried then to confirm the LSND anomaly using a 
beam that excluded the high-energy region of the spectrum,
and it did not see such an excess.

Gninenko has proposed that the LSND anomaly 
could be explained by  a 
50 MeV neutrino, a mass which is beyond the reach at KARMEN. 
This hypothesis requires that 
$\nu_h$ decays into $\nu_i \gamma$ with 
$\tau_h\le 10^{-8}$ s and that
the production mechanism is mediated by the $Z$ 
boson (or other massive particle), as the 
$q^2$ in the collision 
must be large enough to produce the free neutron 
also detected at LSND.

A similar rate of anomalous 
interactions ($3\times 10^{-3}$ per $\nu_\mu$ CC event)
has also been observed at
MiniBooNE both in the neutrino and the antineutrino modes.
In contrast with LSND, 
the energy distribution of these events does
not seem consistent with a simple 3+1 oscillation scheme, 
as it peaks at low energy
(where the $\nu_\mu\to \nu_e$ probability vanishes) and is 
non-significant at the expected oscillation maximum.
The initial neutrino data were actually used
by MiniBooNE to exclude the oscillation hypothesis favored
by LSND. Although the first MiniBooNE 
results in the antineutrino mode were {\it different} and
could favor a 3+1 scheme, the increased statistics obtained
during the 2011 run (in Fig.~7) shows consistency with the 
observations in the neutrino mode (in Fig.~6--right).

\begin{figure}[!t]
\begin{center}
\includegraphics[width=0.48\linewidth]{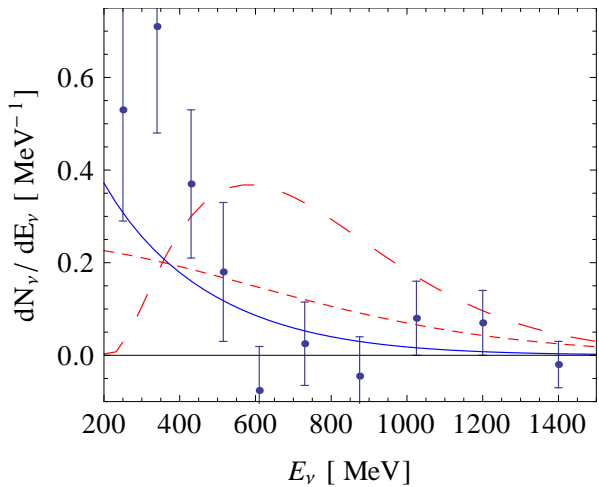}   
\end{center}
\caption{
Energy distribution of 
$\nu_h$ events in the Gninenko model
($|U_{\mu h}|^2= 0.007$, $\tau_h=10^{-9}$ s, $\mu^\mu_{\rm tr}=0$)
(dashes), 
in our set up ($|U_{\mu h}|^2= 0.003$, $\tau_h=5\times 10^{-9}$ s, 
$\mu^\mu_{\rm tr}=2.4\times 10^{-9}\mu_B$) 
(solid),
distribution of events from neutrino oscillations for 
$\sin^2(2\theta)=0.004$ and $\Delta m^2=1\;{\rm eV}^2$ (long dashes), 
and excess observed by MiniBooNE in the neutrino mode.
\label{fig8}} 
\end{figure}
Gninenko's hypothesis can  
have an impact at MiniBooNE only if the
lifetime of the heavy neutrino is reduced below 
$10^{-9}$ s (versus the maximum $10^{-8}$ s required at LSND).
The energy distribution that it implies is {\it better} 
than the one obtained from oscillations, but 
probably still a bit 
flatter than the one observed at MiniBooNE. In Fig.~8 we 
plot 
Gninenko's model for a Majorana $\nu_h$ with 
$|U_{\mu h}|^2= 0.007$ and $\tau_h=10^{-9}$ s.
If $\nu_h$ were a Dirac particle 
its distribution would be even flatter, 
as $\nu_h$ has  then
negative helicity and decays predominantly into a 
forward photon.
At any rate, the mixing and the lifetime in 
Gninenko's model implies too many events with 
radiative muon capture at TRIUMF \cite{McKeen:2010rx}.

Our initial observation is that  the 
coupling required to explain the decay of $\nu_h$
may also imply its production through photon exchange 
with matter. While the lifetime is a measure of 
$\mu^2_{\rm tr}=\sum_i (\mu^i_{\rm tr})^2$, where $i$ runs
over all the light neutrino flavors, the production 
rate at MiniBooNE depends on $\mu^\mu_{\rm tr}$ only.
The photon-mediated production of $\nu_h$ 
will not contribute significantly to the anomaly at LSND,
as the interaction is unable to free a neutron from its
nucleus (it involves charged particles 
and is very soft).

Therefore, we keep the mixing $|U_{\mu h}|^2\approx 0.003$, 
required
to explain LSND through $Z$-mediated interactions, but
we also increase the lifetime $\tau_h$ by a factor of 3--7.
First of all, this reduces in the same proportion 
the number of CC events with 
$\nu_h$ decaying inside the detector at TRIUMF, providing
for consistency with this experiment.
Second, the number of $\nu_h$ decays at MiniBooNE 
will also be reduced,
but this can be compensated by the larger number of 
heavy neutrinos obtained from photon-mediated processes.
Finally, the longer lifetime implies that low-energy 
neutrinos are more likely to decay inside the 
MiniBooNE detector
than the more energetic ones. In addition, 
the negative helicity of the
initial $\nu_\mu$ implies that the EM transition produces a 
(Dirac) $\nu_h$ polarized forward, with positive helicity, and
that the final photon  is emitted preferably backwards.
These two factors define a spectrum 
that seems consistent with the observations at MiniBooNE.

The scenario 
implies that $\nu_h\to \nu_\mu \gamma$ accounts
for just around $1\%$ of the decays, and T2K forces that
the branching ratio for $\nu_h\to \nu_\tau \gamma$ is even
smaller. In any case, an increased statistics at T2K 
could be used to probe the model. 
Other experiments could also put important constraints. Most
notably, NOMAD \cite{Kullenberg:2011rd} has recently 
set bounds (not discussed in previous sections) 
on single photon events in
neutrino interactions at $\approx 25$ GeV.
Although Gninenko \cite{Gninenko:2012rw}
has shown the consistency of his model with the
data, our scenario involves larger lifetimes and 
production cross sections for the heavy heavy neutrino. 
The increase in $\tau_h$ will introduce two competing 
effects that tend to cancell each other: the 
heavy neutrinos produced in the
rock are less likely to decay inside the detector, but they
can be produced further from the detector
and still reach it. We estimate
that for $\tau_h=5\times 10^{-9}$ s our model implies
around $2\times 10^{-4}$ events per each $\nu_\mu$ CC event (above
the $1.6\times 10^{-4}$ limit established there). While lower
values $\tau_h\approx 3\times 10^{-9}$ s would be {\it safer}, 
a more definite statement
would require a full MonteCarlo simulation.

Since the dominant decay channel 
$\nu_h\to \nu_i \gamma$ is not into the
muon nor the tau flavors and the electron flavor seems also
disfavoured, an interesting
possibility would be that the heavy neutrino decays into 
a photon plus {\it another} sterile heavy neutrino $\nu_{h'}$.
The mixing of this second neutrino with the standard flavors
could be much smaller than the one required for $\nu_h$
($|U_{\mu h}|^2= 0.003$), while a  $5$--$10$ MeV  mass
and a lifetime below 1 s would make it consistent with 
bounds from astrophysics and cosmology \cite{Dolgov:2000pj}. 
It could decay $\nu_{h'}\to 
\gamma \nu_{e,\mu,\tau}$ with $c\tau_{h'}> 10$ km. Such a long
lifetime would make it very frequent in the 
atmosphere \cite{Masip:2011qb}:
it would be produced in $\nu_h\to \nu_{h'} \gamma$, 
with the parent $\nu_h$ appearing in 
$\approx 0.3\%$ of all kaon and muon decays \cite{Gninenko:2010pr}.
Such neutrino could also have a relatively large EM dipole
moment,
able to mediate elastic interactions 
$\nu_{h'} Z\to \nu_{h'} Z$
with a cross
section softer but orders of magnitude larger than the ones 
mediated by the $Z$ boson. In particular, at 1--100 MeV energies
one could expect
collisions with 
the typical recoils (few keV) in dark matter experiments.
We think that it would be interesting to analyze to what extent
such neutrino could {\it contaminate} experiments like 
DAMA/LIBRA \cite{Bernabei:2008yi} or CoGeNT \cite{Aalseth:2010vx}.
A similar effect caused by solar neutrinos oscillating into a sterile
flavor has been discussed in \cite{Harnik:2012ni,Pospelov:2012gm}

To conclude, although 
the current framework for neutrino masses and mixings 
explains most of the
experimental data, the picture that we have today is
still far from complete. The possibility
that neutrinos involve new physics at the MeV, a
scale that we have been exploring for many decades, may sound
unlikely or unappealing. We think, however, that it
should be considered, specially as long as 
the current experimental anomalies persist.

\section*{Acknowledgments}
We would like to thank Antonio Bueno, Claudio Giganti and
Patricia S\'anchez-Lucas for discussions.
This work has been partially supported by
MICINN of Spain (FPA2010-16802 and 
Consolider-Ingenio 
{\bf Multidark} CSD2009-00064 and {\bf CPAN} CSD2007-00042),   
by Junta de Andaluc\'{\i}a
(FQM 101, FQM 437 and FQM 3048),
by DFG of Germany (Collaborative Research Center 
{\it The Low-Energy Frontier of the Standard Model}, SFB 1044)
and by MIUR of Italy (Program {\it Futuro in
Ricerca} 2010, RBFR10O36O).

\end{document}